# Emerging Patterns of GenAI Use in K–12 Science and Mathematics Education

## 22% of Math and Science Teachers Report Using GenAI Tools Weekly in Their Teaching

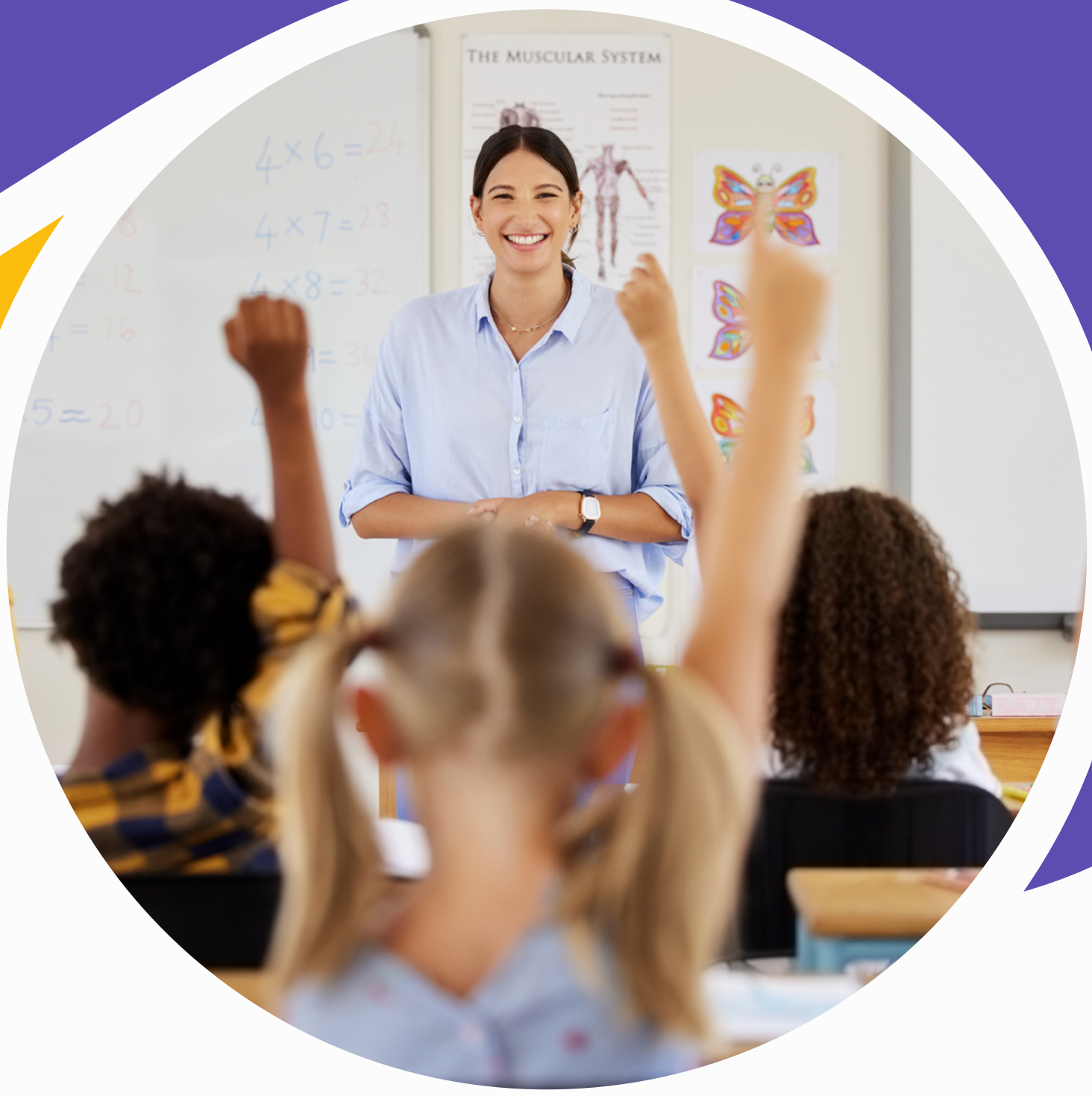

Lief Esbenshade[1], Shawon Sarkar[1], Drew Nucci[2], Ann Edwards[2], Sarah Nielsen[2], Joshua M. Rosenberg[3], Alex Liu[1], Zewei (Victor) Tian[1], Min Sun[1], Zachary Zhang[4], Thomas Han[4], Yulia Lapicus[4], Kevin He[4]

[1] University of Washington, [2] WestEd, [3] University of Tennessee, Knoxville, [4] Colleague AI

AmplifyLearn.AI's work has been supported by the Institute of Education Sciences, U.S. Department of Education, through Grant R305C240012, and by several awards from the National Science Foundation (NSF #2043613, 2300291, 2405110) to the University of Washington and NSF SBIR/STTR award to Hensun Innovation LLC (#2423365). The opinions expressed are those of the authors and do not represent views of the funders.

# Table of Contents

# Executive Summary

In this report, we share findings from a nationally representative survey of US public school math and science teachers, examining current generative AI (GenAI) use, perceptions, constraints, and institutional support. We show trends in math and science teacher adoption of GenAI, including frequency and purpose of use. We describe how teachers use GenAI with students and their beliefs about GenAI's impact on student learning. We share teachers' reporting on the school and district support they are receiving for GenAI learning and implementation, and the support they would like schools and districts to provide, and close with implications for policy, practice, and research. Given the rapid pace of GenAI development and growing pressure on schools to integrate emerging technologies, these findings offer timely insights into how frontline educators are navigating this shift in practice.

## Key Findings

- As of spring 2025, over 50% of surveyed Math and Science teachers reported that they have used GenAI tools in their teaching. 42% of teachers who used GenAI tools reported using them at least weekly.
- Teachers who used GenAI primarily used it for instructional planning and preparation (76%) and creating student assessments (62%), and only infrequently for automating grading of student work (only 13%).
- Surveyed teachers - including those who did not use GenAI - were roughly split into thirds regarding the impact on student learning. 34% believed that GenAI has had a negative impact, 36% found no effect, and 30% believed the effect has been positive. 40% of teachers believed that students misuse GenAI for plagiarism or copying answers.
- Teachers reported limited district guidance: only 5% reported that their district has formal guidelines in place, and only 22% reported receiving any district training on GenAI.
- 75% of teachers wanted professional training on the GenAI tool basics and capabilities. More than two-thirds wanted professional development on lesson planning, differentiation, and assessment applications of GenAI, and 65% wanted professional learning that helps them guide students in their GenAI use.

# Survey Implications

Our study found that **50% of math and science teachers used GenAI in their teaching, with 22% reporting using GenAI at least weekly.** These findings complement other recent national research on GenAI adoption in K-12 math and science education. Gallup & Walton Family Foundation (2025) surveyed 2,232 K-12 public school teachers across disciplines and found that 59% of math teachers and 66% of science teachers used GenAI tools during the 2024-25 school year. Compared to a prior RAND survey in the 2023-24 school year that found only 19% of STEM teachers used GenAI in their work (Dilberti, et al., 2024) - these findings suggest that GenAI use more than doubled from the 23-24 through 24-25 school year.

**Math and science teachers reported using GenAI tools predominantly for lesson planning, creating assessments, and providing student feedback** rather than using GenAI during classroom instruction. These findings indicate that math and science teachers are strategically leveraging AI to enhance their pedagogical preparation while maintaining their essential role as content experts and instructional leaders in the classroom.

Nearly half of GenAI-using teachers reported that GenAI tools are changing their student interactions, primarily through enhanced personalization and differentiated instruction. However, teachers had mixed reviews of the impact of GenAI on student learning. While **30% of surveyed teachers said the use of GenAI positively impacted their students' learning, 34% said GenAI had no impact on student learning,** and 40% of teachers cited student misuse of GenAI and cheating as major concerns.

Teachers report needing more time and more training opportunities to learn how to use GenAI tools effectively. While Diliberti and colleagues found that 48% of districts had provided some form of GenAI training to teachers by Fall 2024 (Diliberti et al., 2025), in our study, **only 22% of math and science teachers reported receiving district training, even though 75% indicated a desire for training on GenAI basics specifically.** These findings highlight a critical gap between educators interest in GenAI learning and institutional support for it. As GenAI tools become increasingly sophisticated, the 72% of teachers who remain either non-users or infrequent users represent a significant opportunity for targeted professional development that addresses the technical capabilities, pedagogical applications, and ethical considerations of GenAI.

While **86% of GenAI-users employ it to improve efficiency, far fewer—just 20–46%—are using GenAI to enhance instruction** with richer, more relevant learning experiences. Limited district support appears to be a key barrier to GenAI adoption: **only 5% of teachers work in schools with formal GenAI policies,** and just 9% report encouragement to integrate it into instruction. Teachers cite a lack of professional development, unclear policies, and time to learn as the biggest obstacles to using GenAI for instructional tasks, even as 90% rely on ChatGPT, a tool not designed for teaching math and science. The findings point to an urgent need for district-level policies, endorsed instructional tools, and targeted professional learning for teachers and leaders. Without this support, GenAI's potential to improve teaching and learning will remain largely untapped.

## Survey Sample & Administration

RAND fielded the survey to its American Teacher Panel between April 30, 2025, and May 25, 2025. This panel consisted of 25,000 educators. To learn more information about the panels, visit https://www.rand.org/education-and-labor/survey-panels.html. The sample included public school teachers who teach math or science in grades K-12 (n = 979). We surveyed 352 secondary mathematics teachers, 163 secondary science teachers, and 464 elementary teachers who teach both science and mathematics. To maintain analytical quality, in the report that follows, we do not disaggregate the data into these subsets. Technical details on survey administration and analysis can be found at the end of this report.

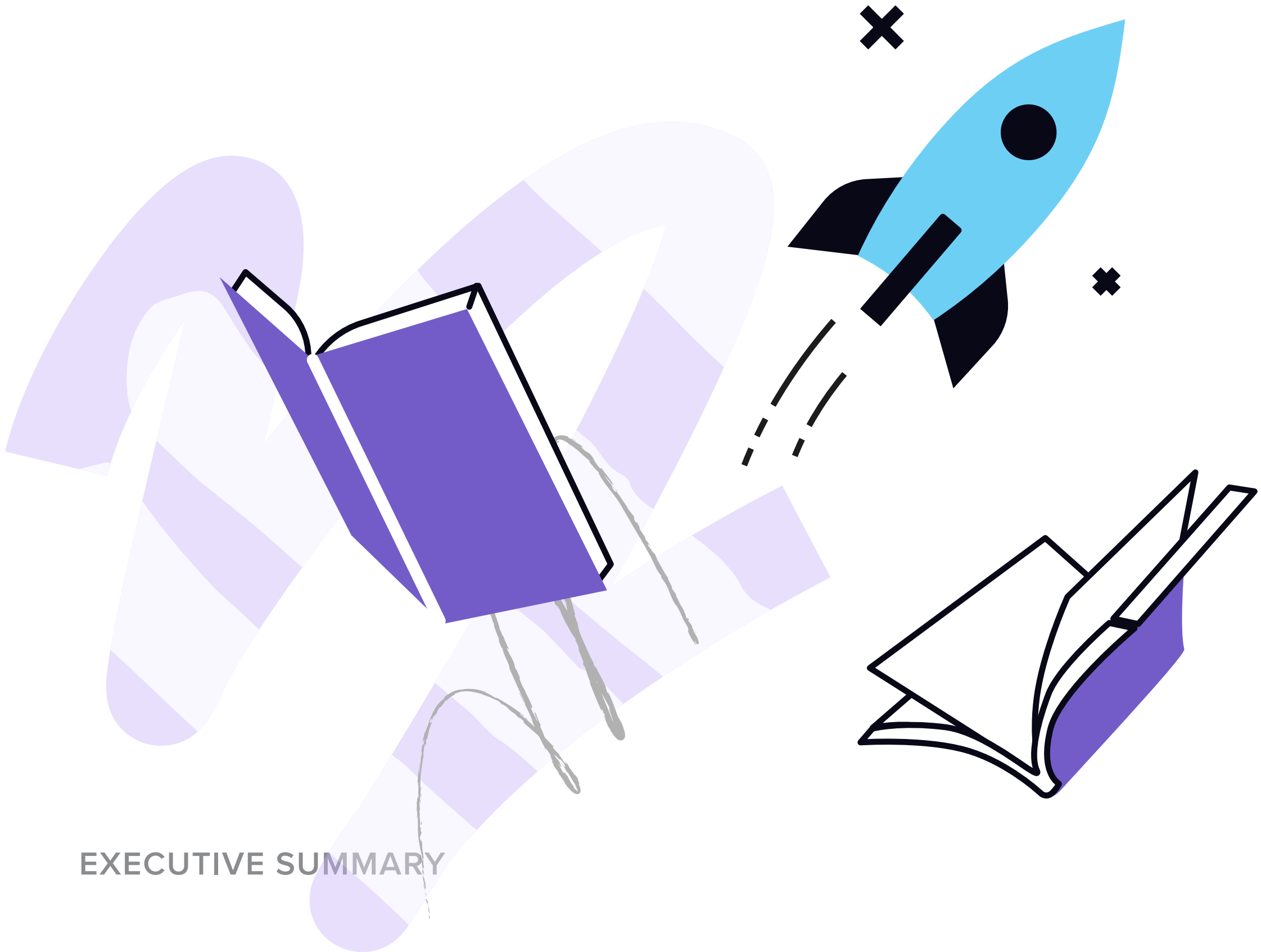

# GenAI Tools Usage in Math and Science Teachers' Educational Practice

## 50% of Teachers Have Used GenAI

Survey findings suggest that teachers' integration of GenAI into instructional tasks is still in the early stages but is potentially growing. While 50% of teachers have used GenAI, over one-third (35%) said they had tried using GenAI and plan to use it more often. A further 10% reported regular or advanced use, and 24% of respondents said they had not yet used GenAI tools but did not choose the option indicating a refusal of future use. 14% of teachers chose not to use GenAI tools, and a further 5% have tried them but do not intend to use them again.

**Which best describes your current use of GenAI tools in teaching?**

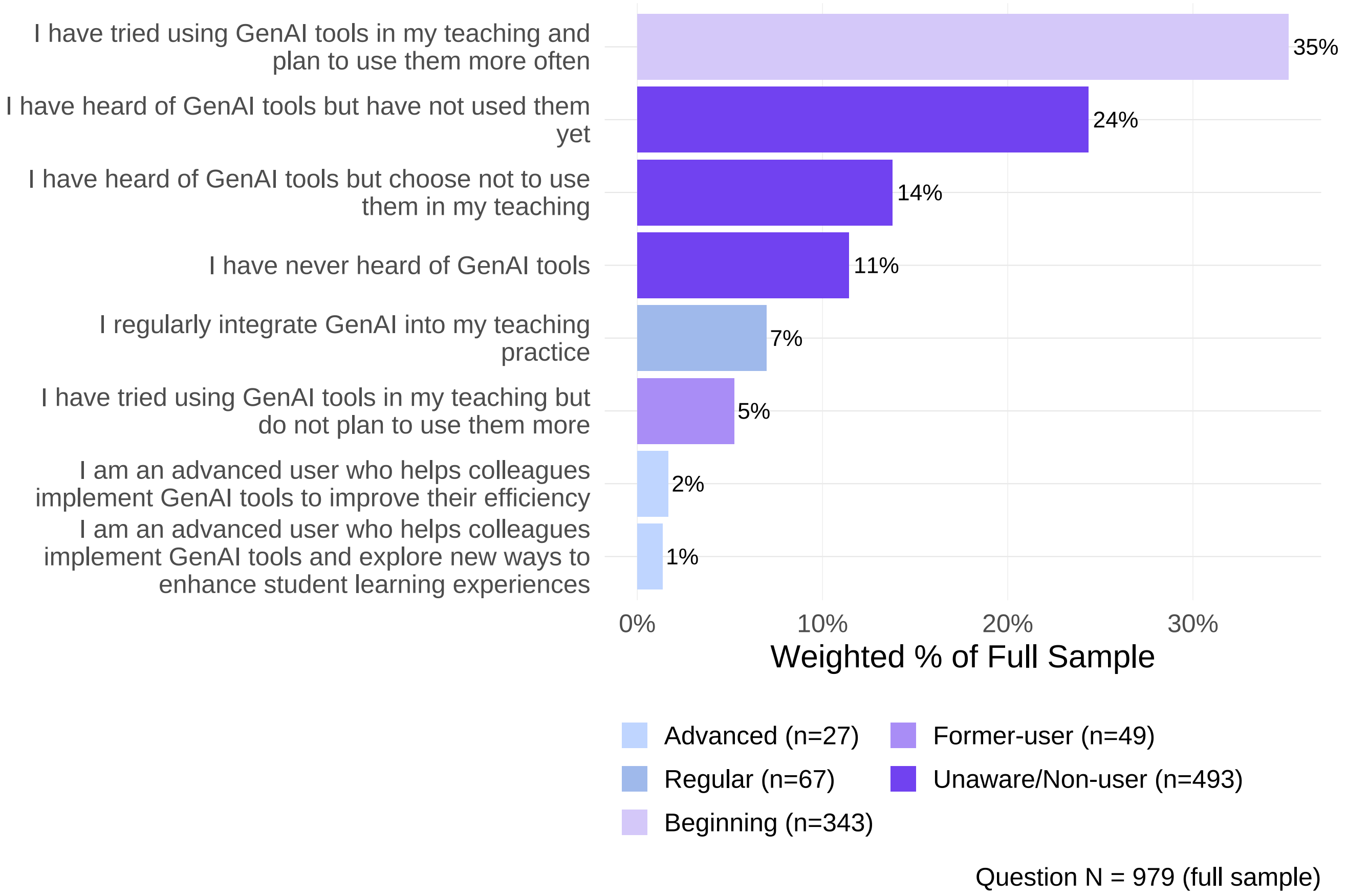

The next several sections pertain to the frequency and purposes of teachers' GenAI use and teachers' understandings of the impact of GenAI on their students. As such, the data presented in these sections only pertains to those who indicated that they had used GenAI tools (n = 486). Percentages are reported for this subset only and should therefore be interpreted as representative of US public school math and science GenAI-using teachers rather than all US public school K-12 math and science teachers.

## Most GenAI Using Teachers Are New and Infrequent Users

GenAI integration into math and science instructional practice is nascent. Of the 50% of teachers who reported using GenAI tools (n=486), most were infrequent users and recent adopters. 57% of GenAI-using teachers reported infrequent use (monthly or rarely) while 42% indicated more frequent use (weekly or daily). The majority (76%) indicated that they have been using GenAI for less than a year. 64% of teachers indicated that they increased their usage of GenAI tools compared to when they first started using them, with only 4% indicating less frequent use.

**On average, how frequently do you use GenAI tools for instructional tasks (lesson planning, teaching, assessment, professional learning, and/or administrative tasks)?**

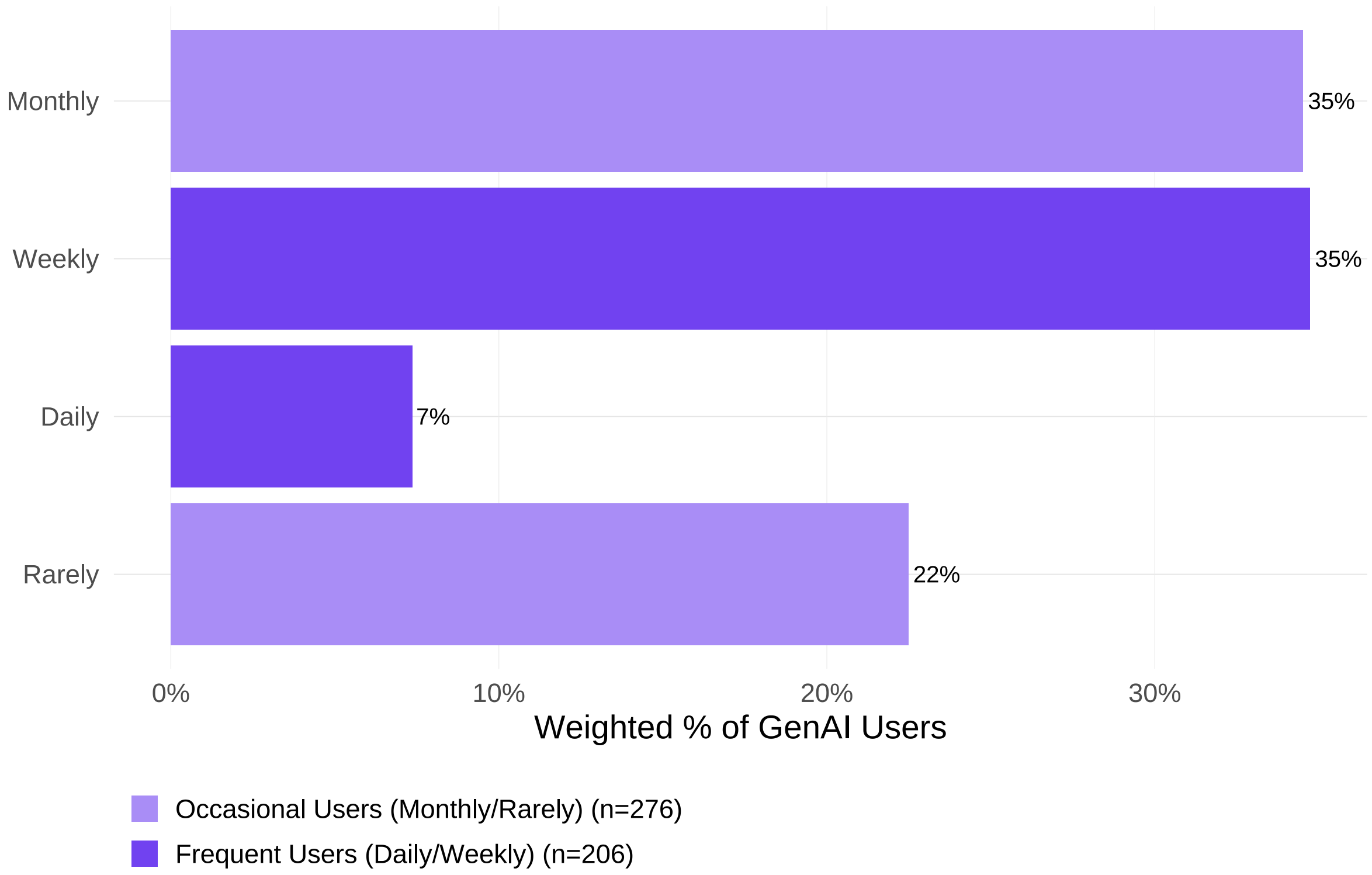

Question N = 486 (genAI users)

## How long have you been using GenAI tools?

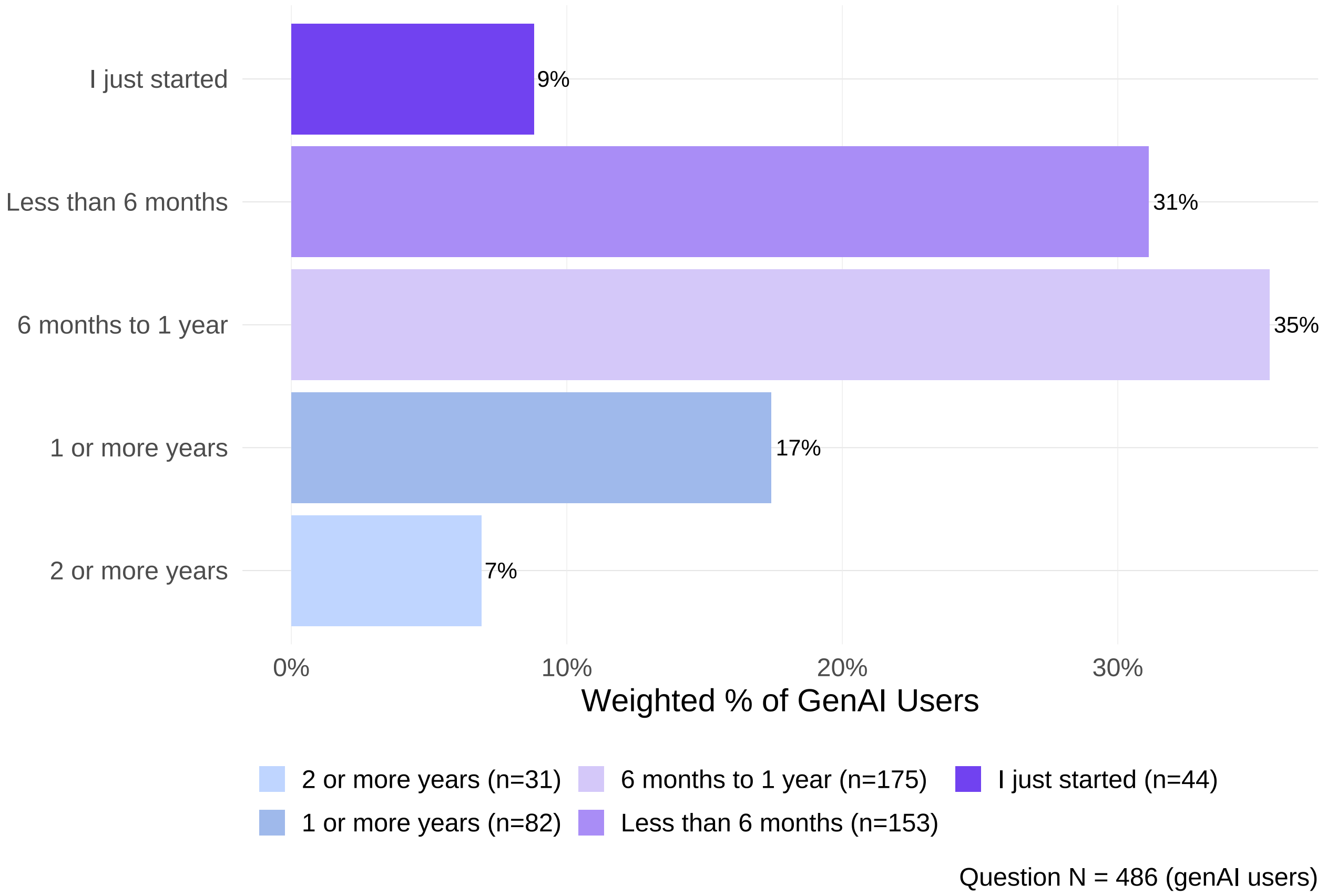

## How has your use of GenAI tools changed since you first started using them?

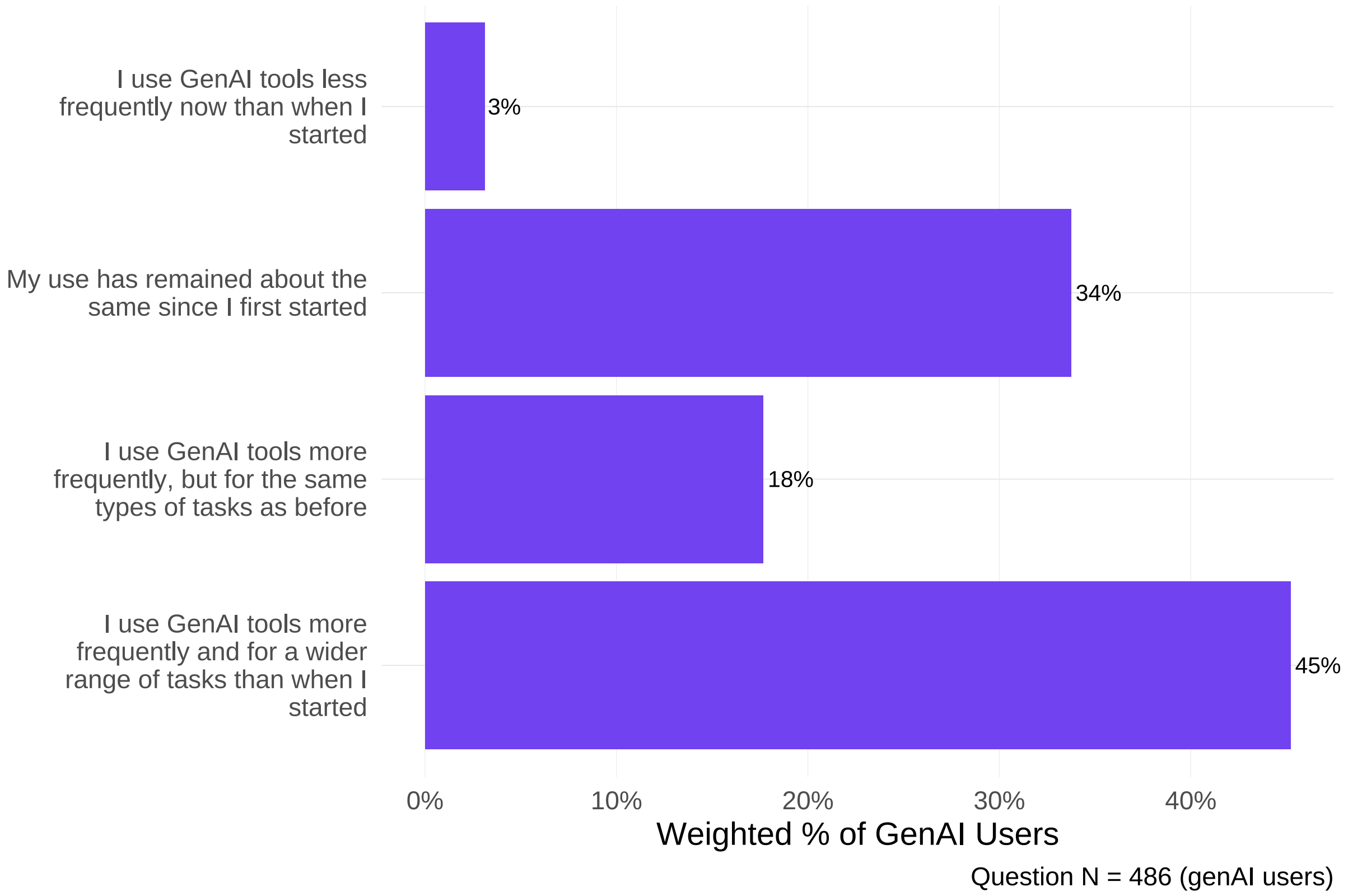

# Teachers Use GenAI to Create Better Instructional Materials

Among the 50% of math and science teachers using GenAI tools (n=486), 76% and 61% reported using GenAI for lesson planning and assessment development, respectively. 50% reported using GenAI in class, with 10% reporting that they taught students about deploying GenAI as a learning tool. Open-ended responses about teacher use cases revealed that many schools have restricted or blocked student access to tools like ChatGPT due to ethical, privacy, and academic integrity concerns, limiting teachers' ability to deploy GenAI for classroom activities. Elementary teachers also noted that their students are too young to engage meaningfully with GenAI tools. Beyond classroom instruction, teachers reported using GenAI for a range of professional tasks, including writing grants, drafting emails to parents, and generating multimedia content for presentations or instructional materials.

### For which instructional tasks have you used GenAI tools?

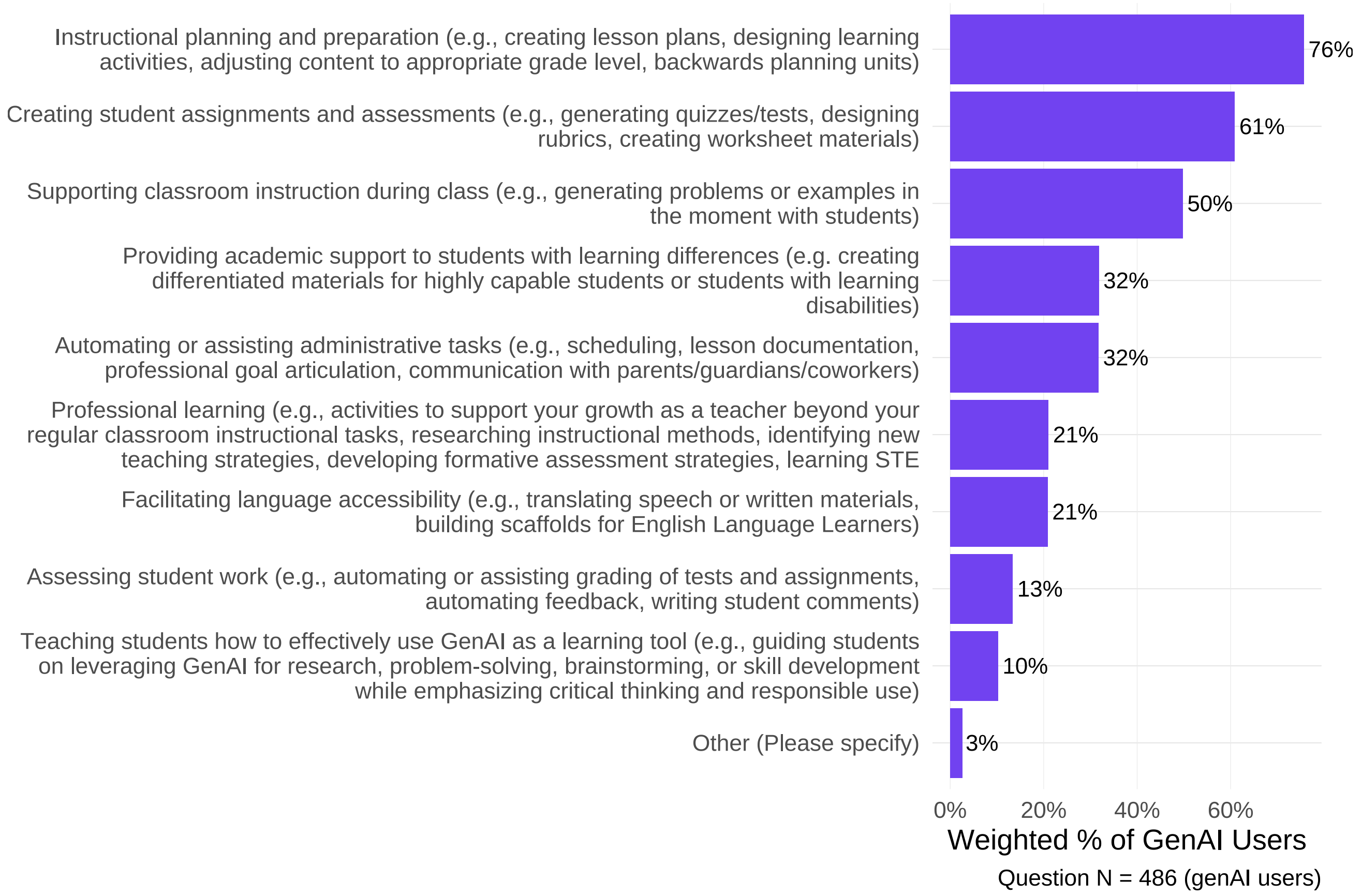

Teachers found GenAI most valuable for content creation and instructional planning, with 46% identifying improved material creation as a key benefit of GenAI. Teachers also reported improvements in other elements of lesson planning, with direct connections to the kinds of learning experiences afforded to students. For example, 43% used GenAI to generate better questions to ask students in class, and 31% for better conceptual explanations. 29% used GenAI to make math and science more relevant to their students, with 27% connecting to real-world examples and 23% creating more dynamic visual representations for students. Notably, few teachers reported GenAI benefits to facilitating in-person interactions like collaboration and student problem-solving.

In addition to lesson planning and instruction, teachers reported some benefits of GenAI for other aspects of the work of teaching. For example, 86% of GenAI-using teachers indicated that GenAI helped them work at least somewhat more efficiently on routine tasks, and 67% indicated GenAI at least somewhat helped them build a more cohesive and standards-aligned curriculum. Only half of the teachers reported improvements in tracking student progress.

**How has your use of GenAI impacted your classroom teaching and approach to educational content specifically?**

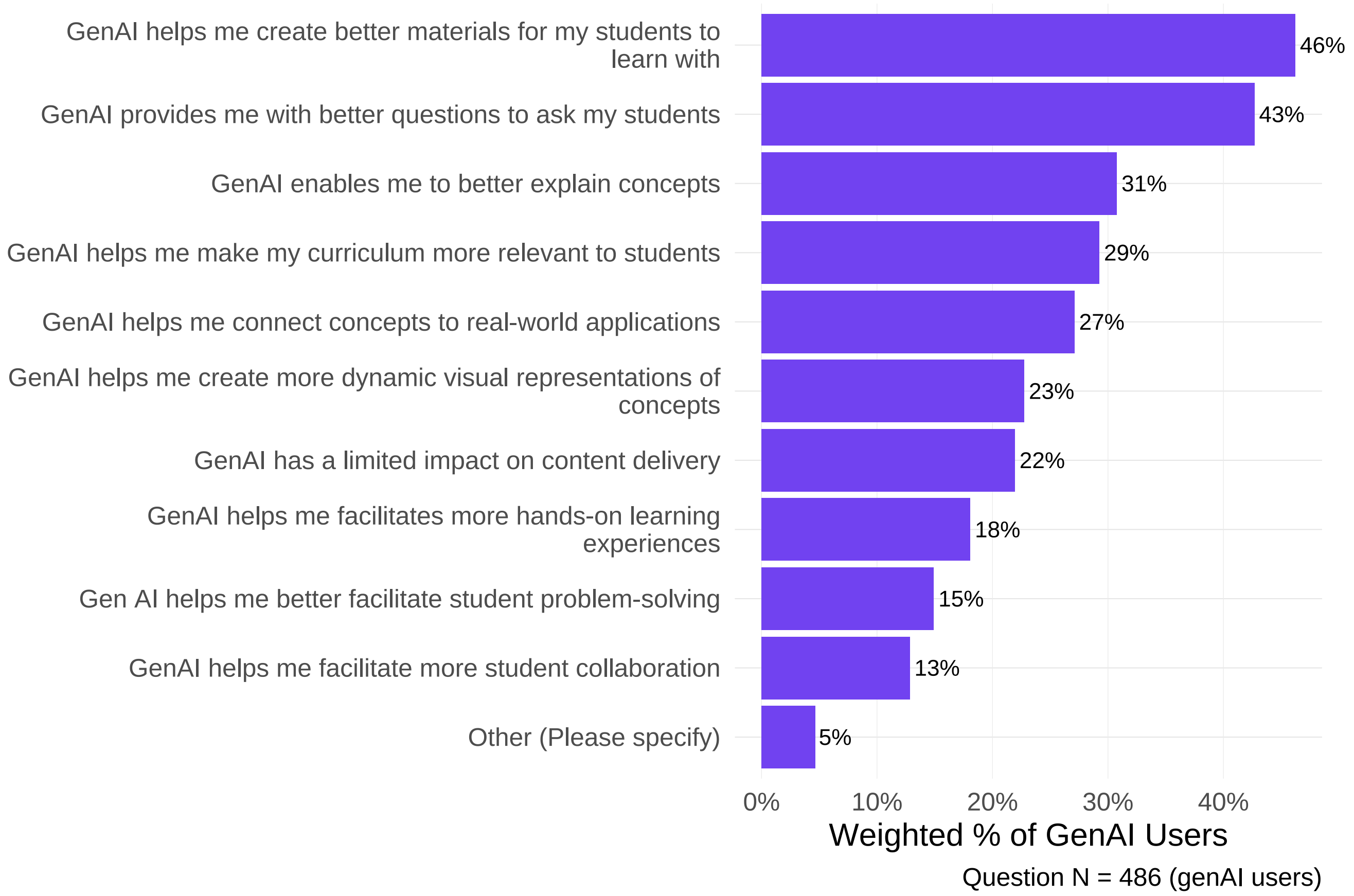

**In your experience, to what extent have GenAI tools impacted your planning, teaching, assessment, professional learning, and/or administrative tasks?**

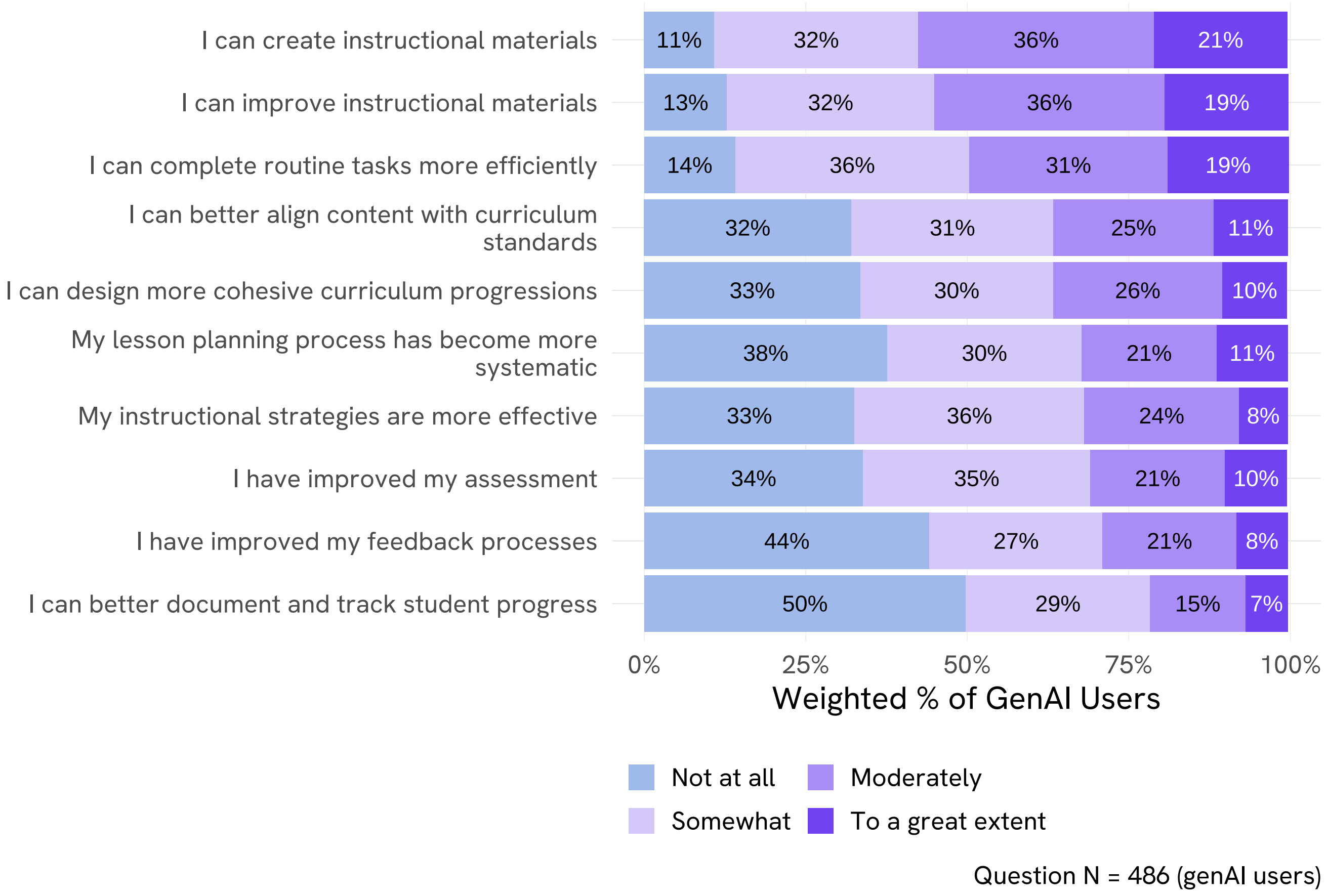

# ChatGPT is the Dominant GenAI Tool Used by Teachers

The ed-tech marketplace remains fragmented, with many new GenAI tools designed for teachers, but ChatGPT was the dominant GenAI tool used by teachers, with 88% of GenAI-using teachers indicating that they had used it. MagicSchool also stood out in our survey responses, with 45% of GenAI-using teachers reporting its use; all other tools were at 22% usage or lower.

**Which GenAI tools have you used for instructional tasks (lesson planning, teaching, assessment, professional learning, and/or administrative tasks)?**

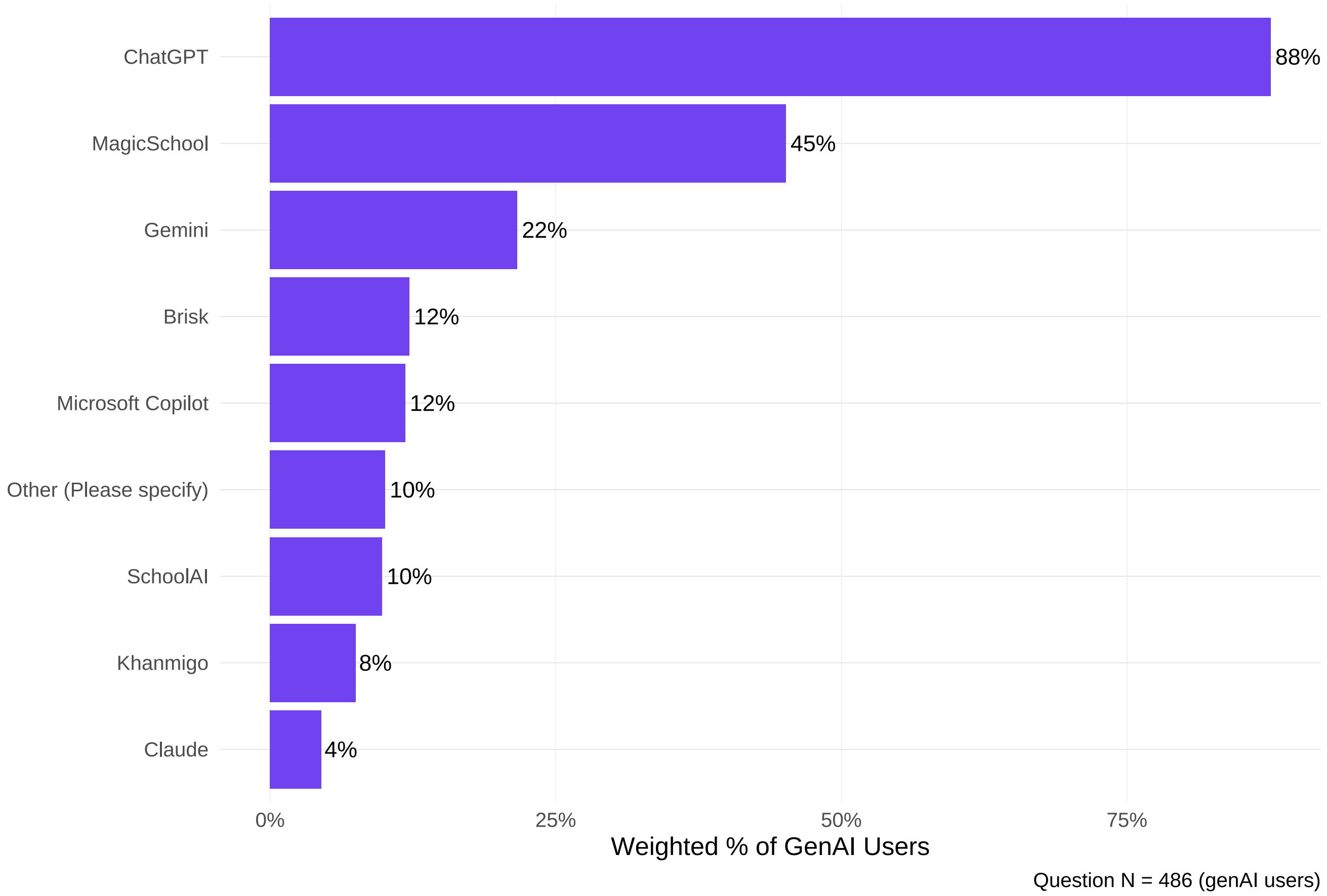

# GenAI and Teacher-Student-Content Interactions

## Some Teachers Report Improved Personalization of Learning with GenAI Tools

47% of the teachers who used GenAI tools reported that GenAI was at least somewhat changing their interactions with students. These teachers (n = 277) reported that creating more personalized learning experiences and better accommodating different learning needs were the top changes they observed.

To an open-response prompt, some teachers commented that GenAI added to their workload, as they had to spend time detecting AI-generated student work and reinforcing the value of independent effort, describing it as a continual struggle to ensure students are doing the intellectual "heavy lifting" themselves.

**To what extent has GenAI changed your interactions with students?**

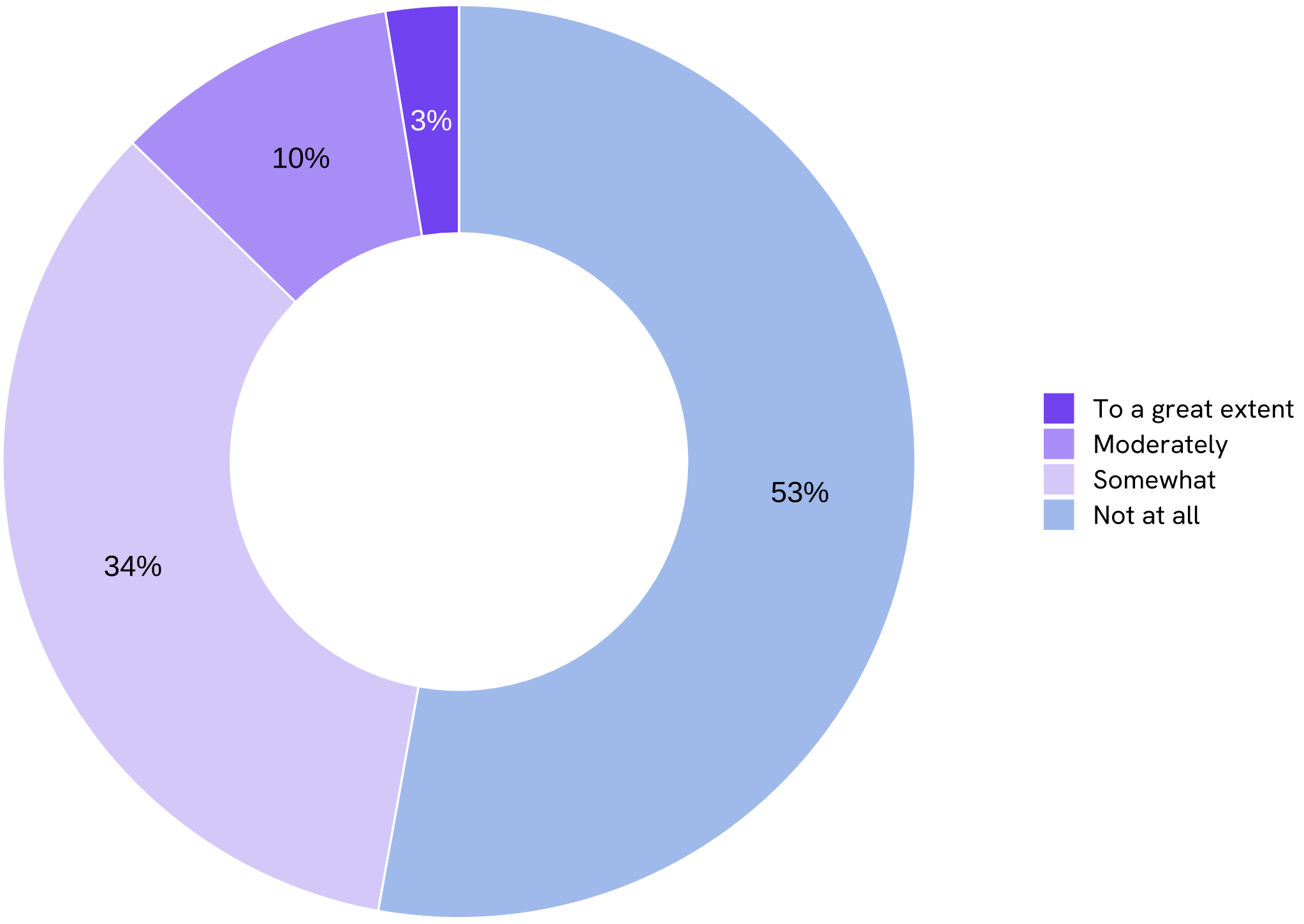

Question N = 486 (genAI users)

## How has your use of GenAI changed your interactions with students?

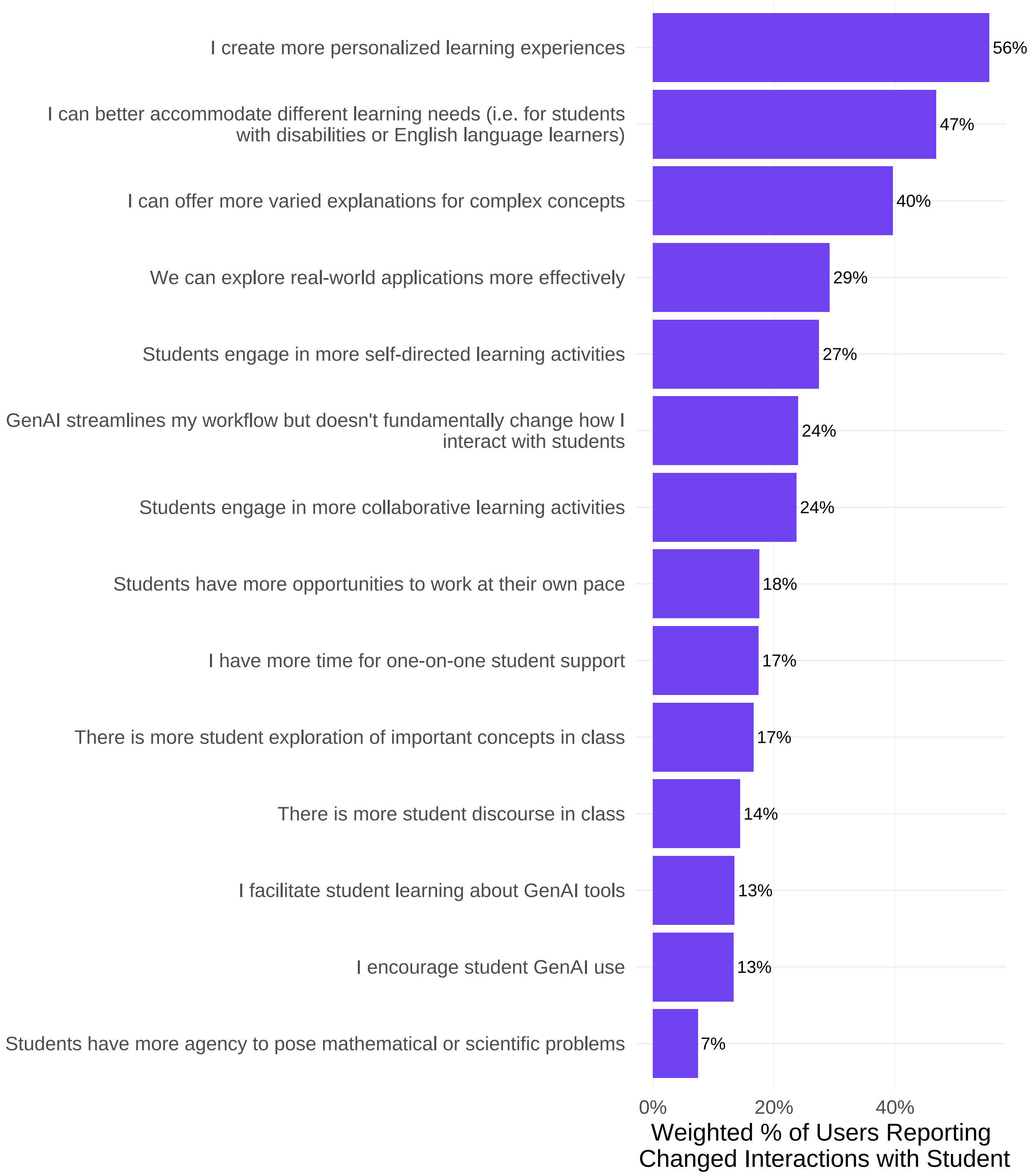

Question N = 227 (Changed interactions with students)

# Teacher Perceptions of Student Use of GenAI Tools

## Teachers Were Split on GenAI's Impact on Student Learning: 34% Negative, 30% Positive

So far in this report, we have been discussing GenAI use patterns - frequency, recency of adoption, purposes, and tools. We return now to the whole dataset (n = 979), which includes both GenAI users and non-users. Survey results reveal mixed educator perspectives on GenAI's impact on student learning, roughly cutting the data in thirds. 30% of teachers viewed the impact of GenAI on students positively, 34% negatively, and the largest group (36%) reported neither positive nor negative impacts. When students do engage with GenAI tools, teachers reported that they use them primarily for homework assistance (21%) and building concept understanding (15%). However, academic integrity concerns dominate the challenges, with 40% of educators reporting plagiarism issues and 35% noting over-reliance on AI for basic tasks. Open-text responses indicate that many schools blocked student access to tools like ChatGPT due to ethical, privacy, and academic integrity concerns, while elementary educators noted their students may be too young for meaningful GenAI engagement. This creates a 'dual use dilemma' where the same technology can both support and undermine learning, highlighting the critical need for structured guidance and AI literacy curricula to help students navigate these tools appropriately.

**How negatively or positively do you think the use of GenAI has impacted students' learning overall?**

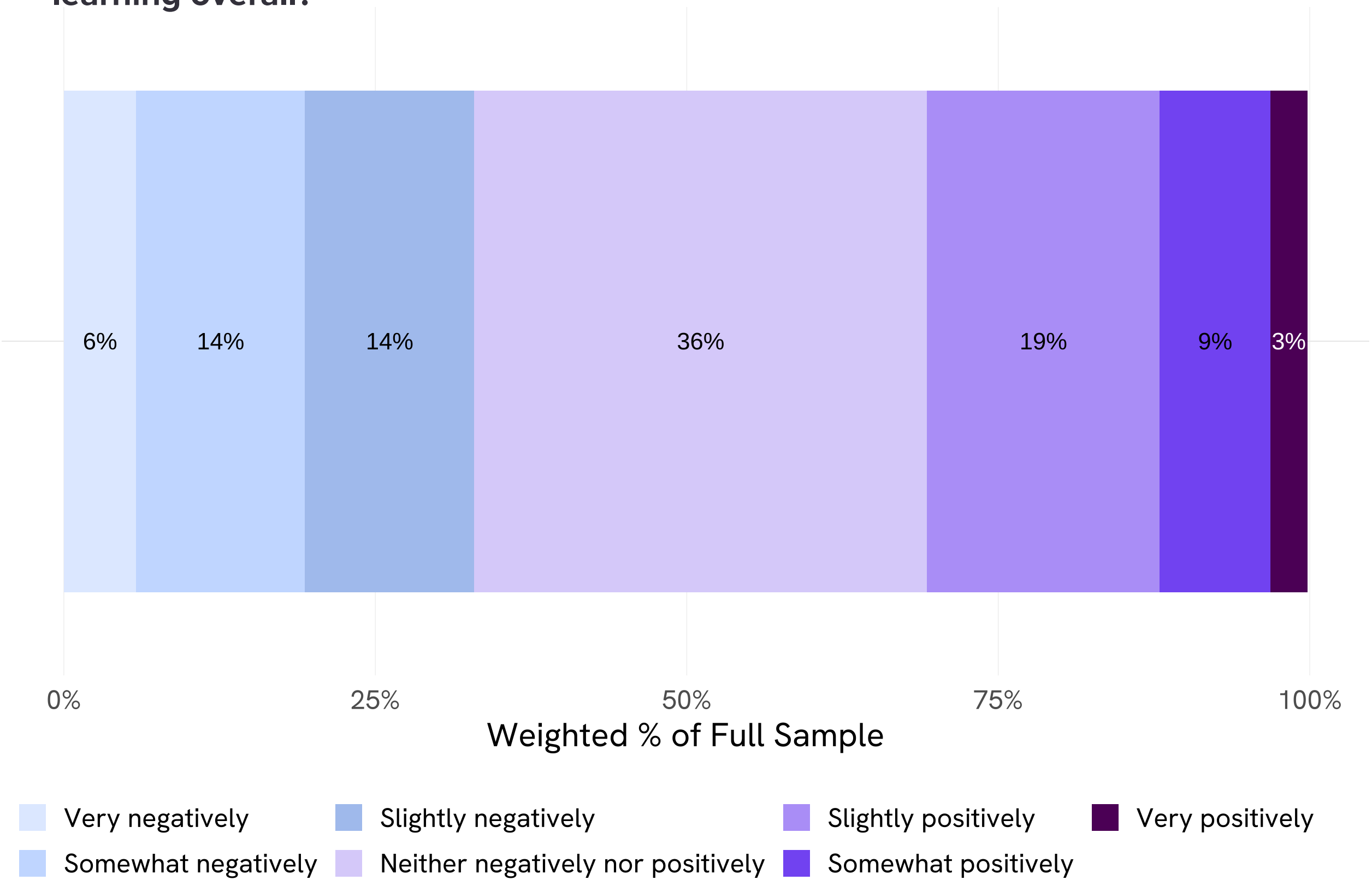

Question N = 979 (full sample)

### How do students interact positively with GenAI tools in your classroom?

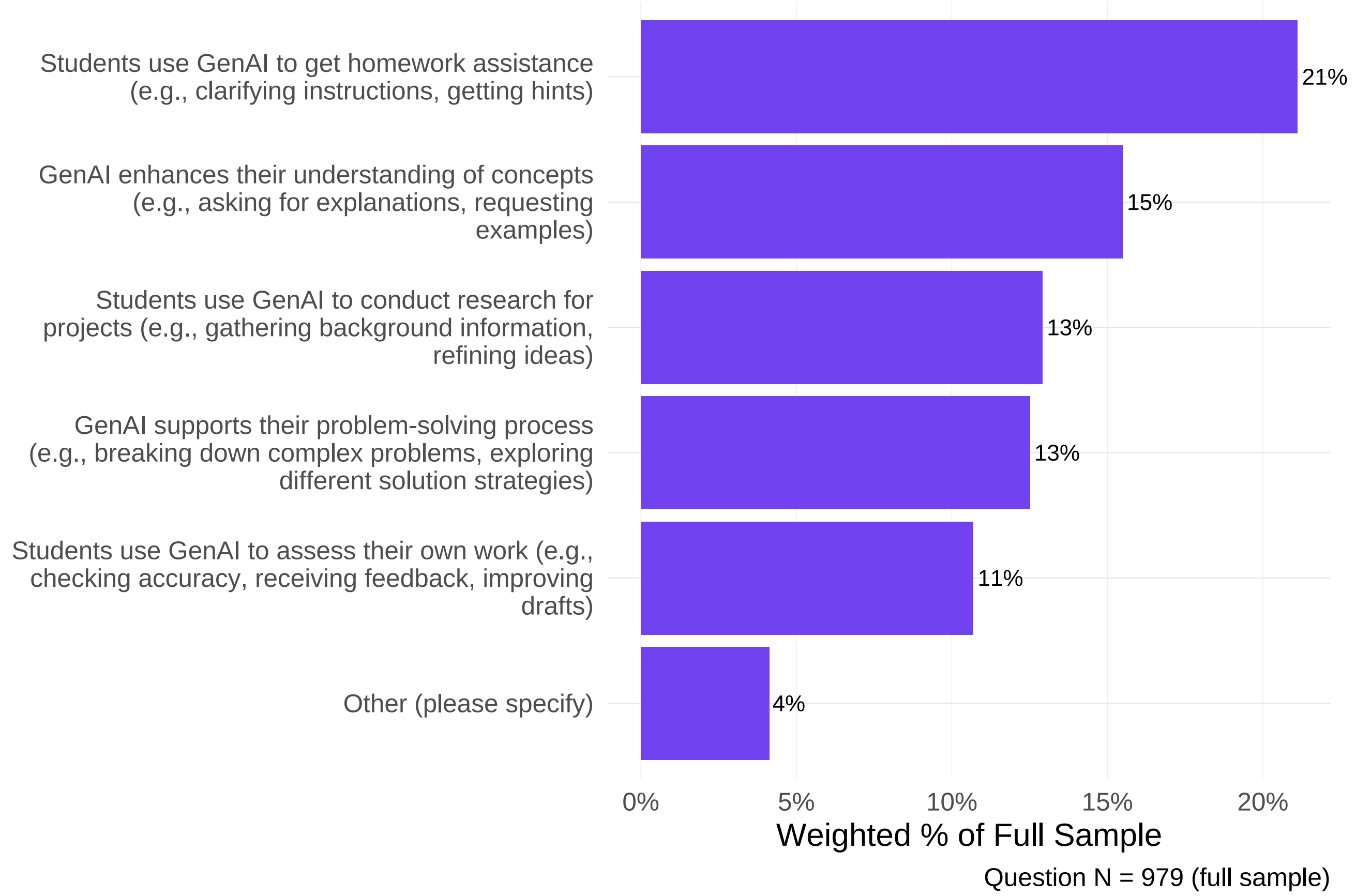

### How do students interact negatively with GenAI tools in your classroom?

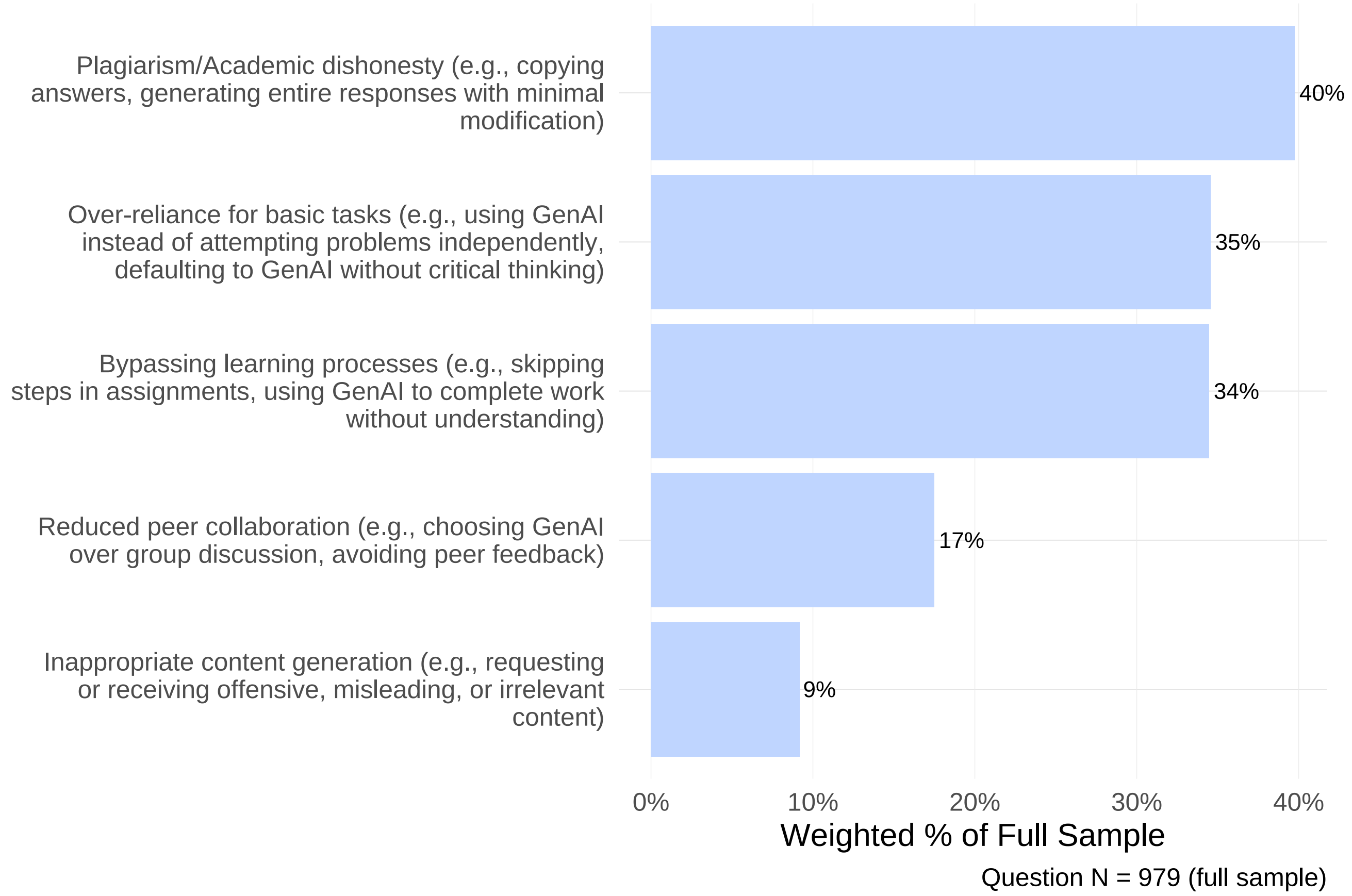

# School Policies and Professional Development

## School Districts are Still Developing Guidelines for GenAI

48% of all surveyed teachers (n = 979) reported that their schools have no GenAI guidelines, with 42% reporting informal policies or policies in development. Among the 52% of teachers reporting that their schools that do have some form of guidelines (n = 495), 41% allow both teacher and student use (with restrictions), and roughly equal proportions fully encouraging integration across subjects (17%), permitting use in specific subjects or grade levels (22%), or limiting use to teachers only (18%).

**What is your district's current approach to GenAI use in education?**

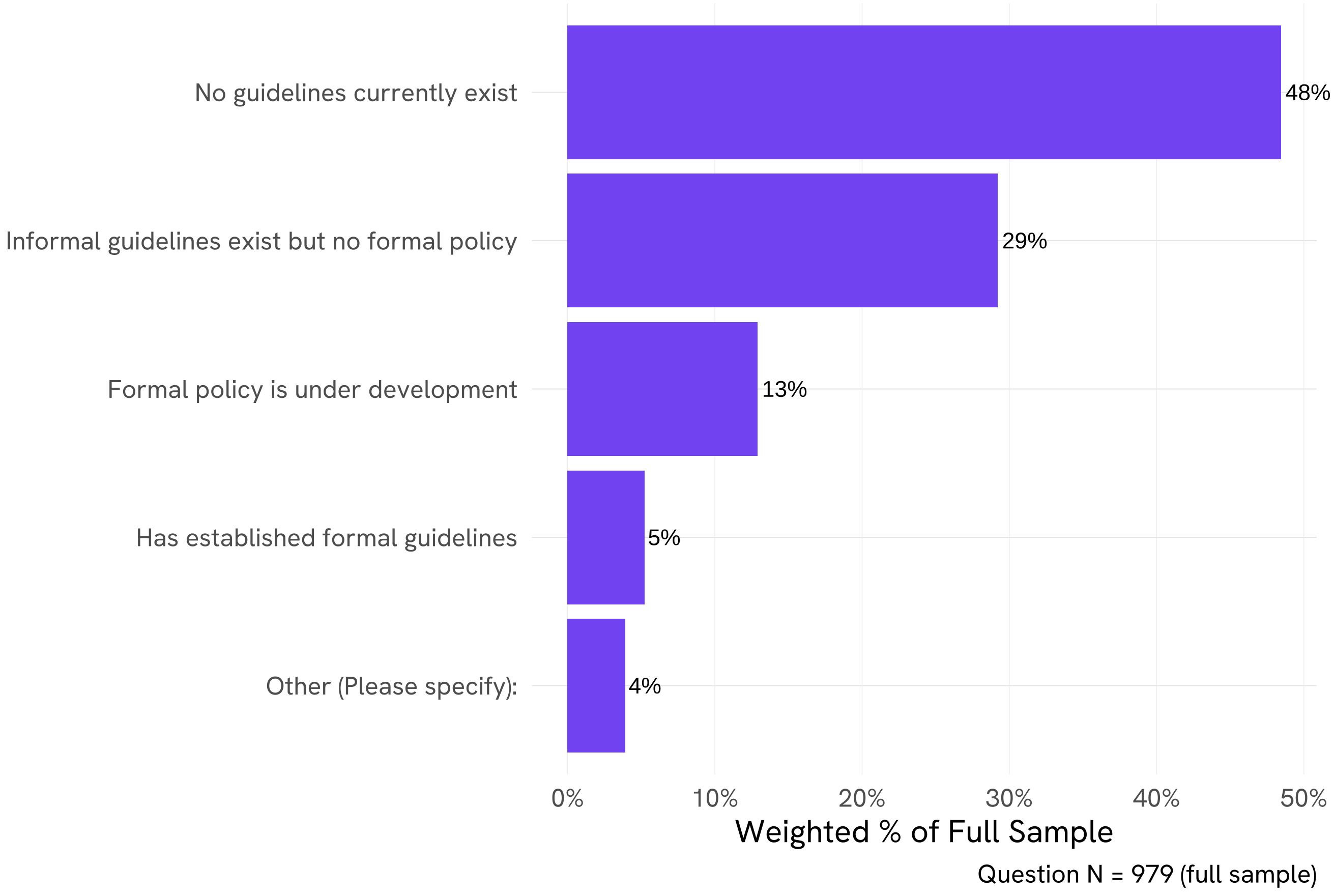

**Which of the following best describes your district's AI guidelines or policies?**

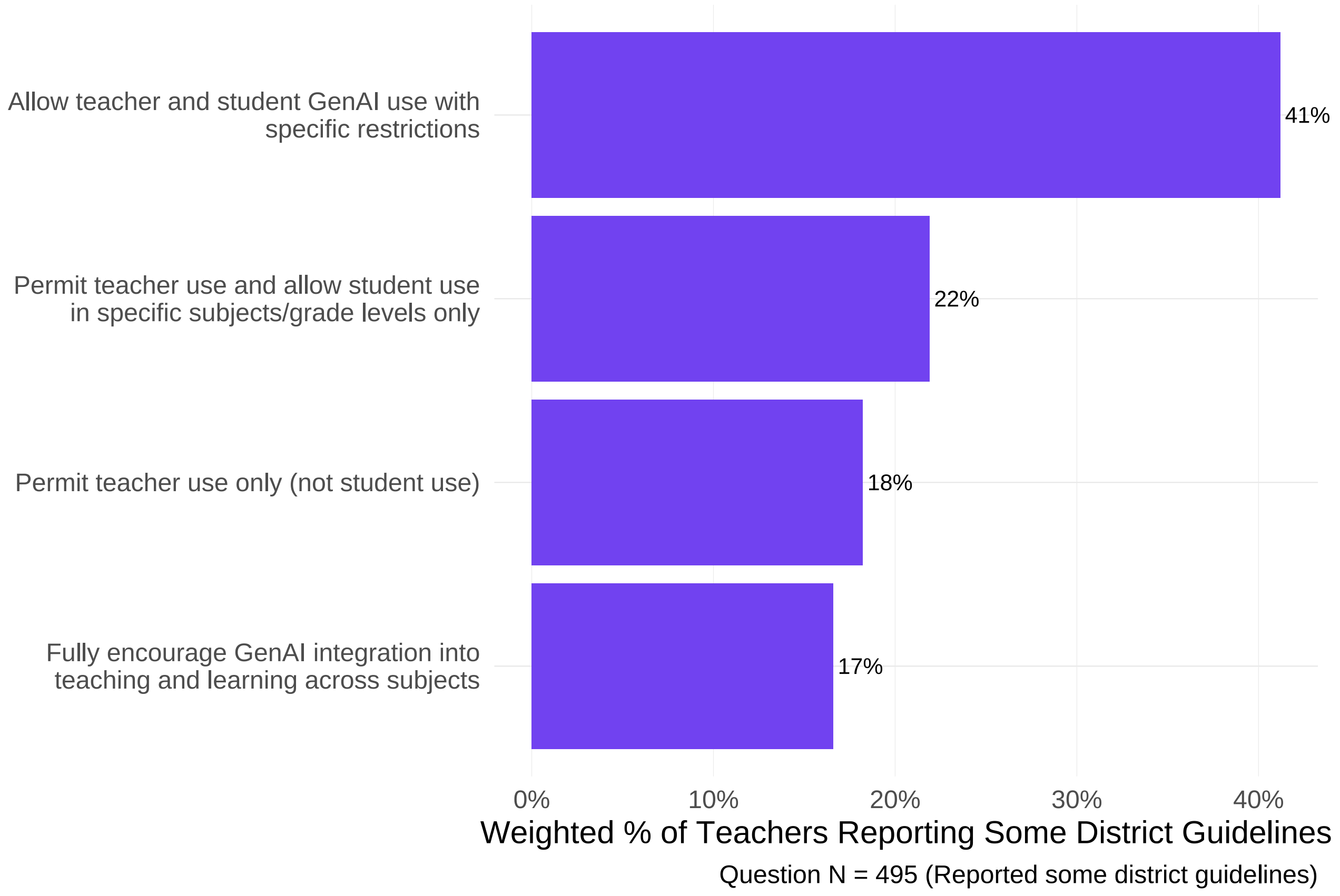

## Teachers Still Need Basic Training on GenAI Use

Survey data reveals a critical professional development deficit. Teachers (n=979) showed overwhelming demand for practical GenAI implementation training, prioritizing tool basics (75%), lesson planning (73%), and differentiation techniques (69%, with lower demand for ethical use guidelines (46%) and data privacy (36%). Vitally, 65% seek professional development on helping students understand and use GenAI critically. Despite the demand, 66% of all teachers surveyed reported receiving no formal GenAI training, with only 22% receiving district sessions and 13% relying on informal colleague learning.

### Select all of the types of professional training you would find helpful to support your GenAI adoption:

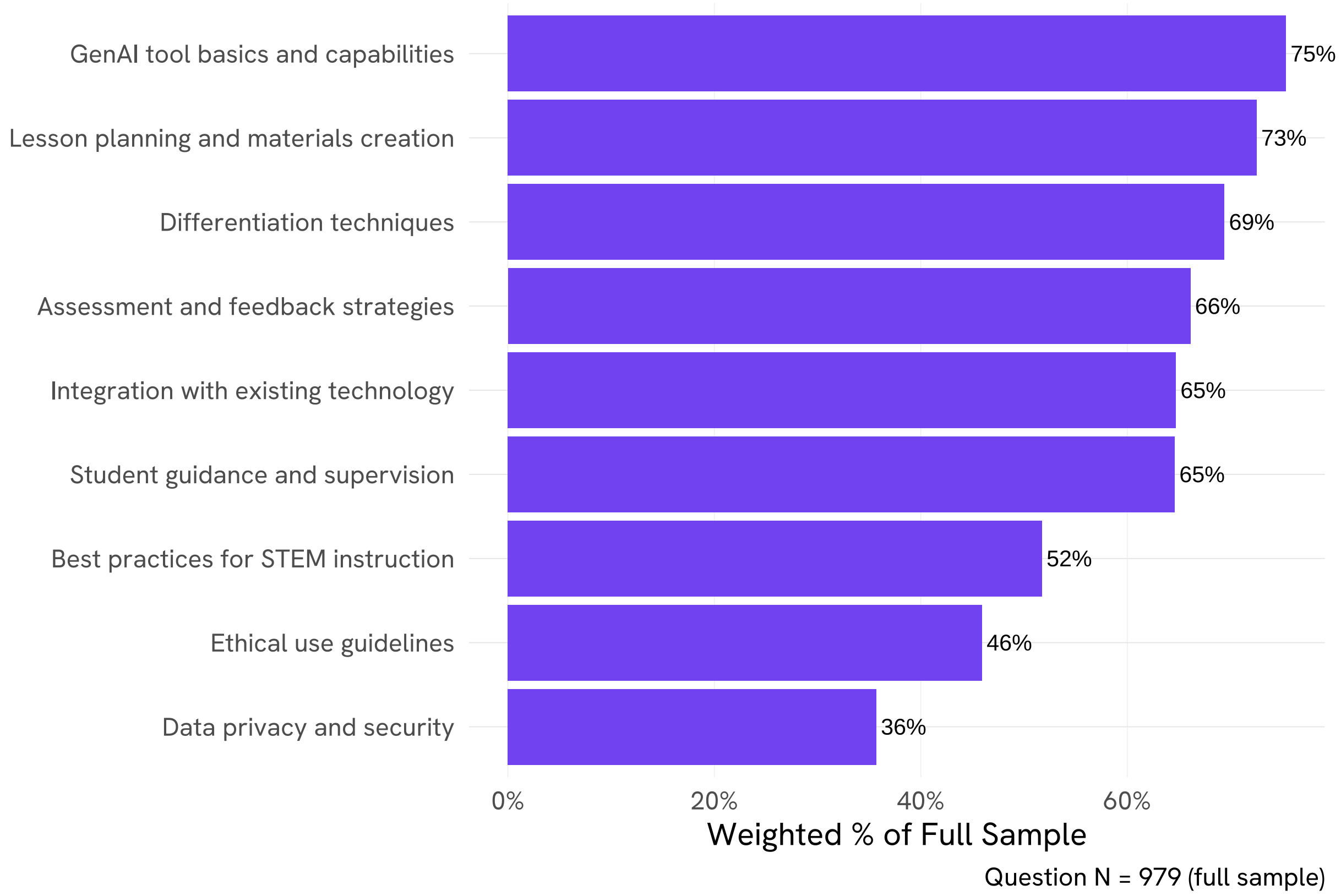

### Have you received any training on using GenAI tools?

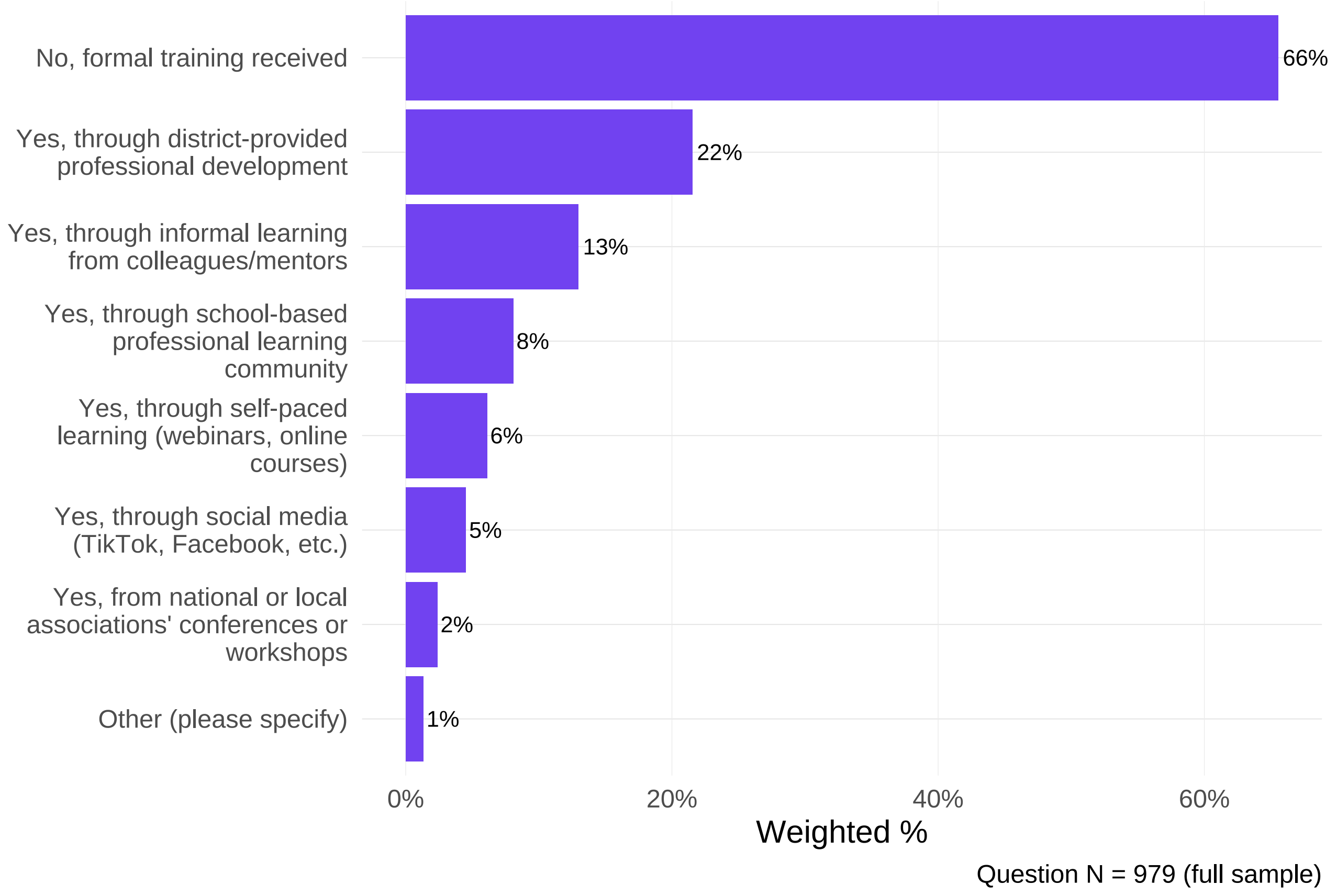

# Challenges and Concerns for Adopting GenAI

## Time, Training Gaps, and Worries About Student Reliance on GenAI

The greatest obstacles to GenAI adoption are practical learning challenges, with 61% of teachers citing the time required to learn these tools as a barrier to adoption. 54% reported insufficient training opportunities, consistent with the finding that two-thirds of teachers have not received formal GenAI training. Additionally, 45% reported struggling with unclear district guidelines. This is perhaps unsurprising given the finding that only 5% of teachers reported well-formed and clear GenAI policy guidelines at their schools. Notably, only 9% cite restrictive policies as barriers, suggesting the challenge for adoption is insufficient institutional support rather than prohibition. Teachers' primary concerns center on student over-reliance (62%) and replacing human creativity (60%), followed by content accuracy issues (54%), but they show relatively low worry about GenAI affecting teacher-student relationships (22%).

> **What challenges do you face in adopting and implementing GenAI for instructional tasks (lesson planning, teaching, assessment, professional learning, and/or administrative tasks)?**

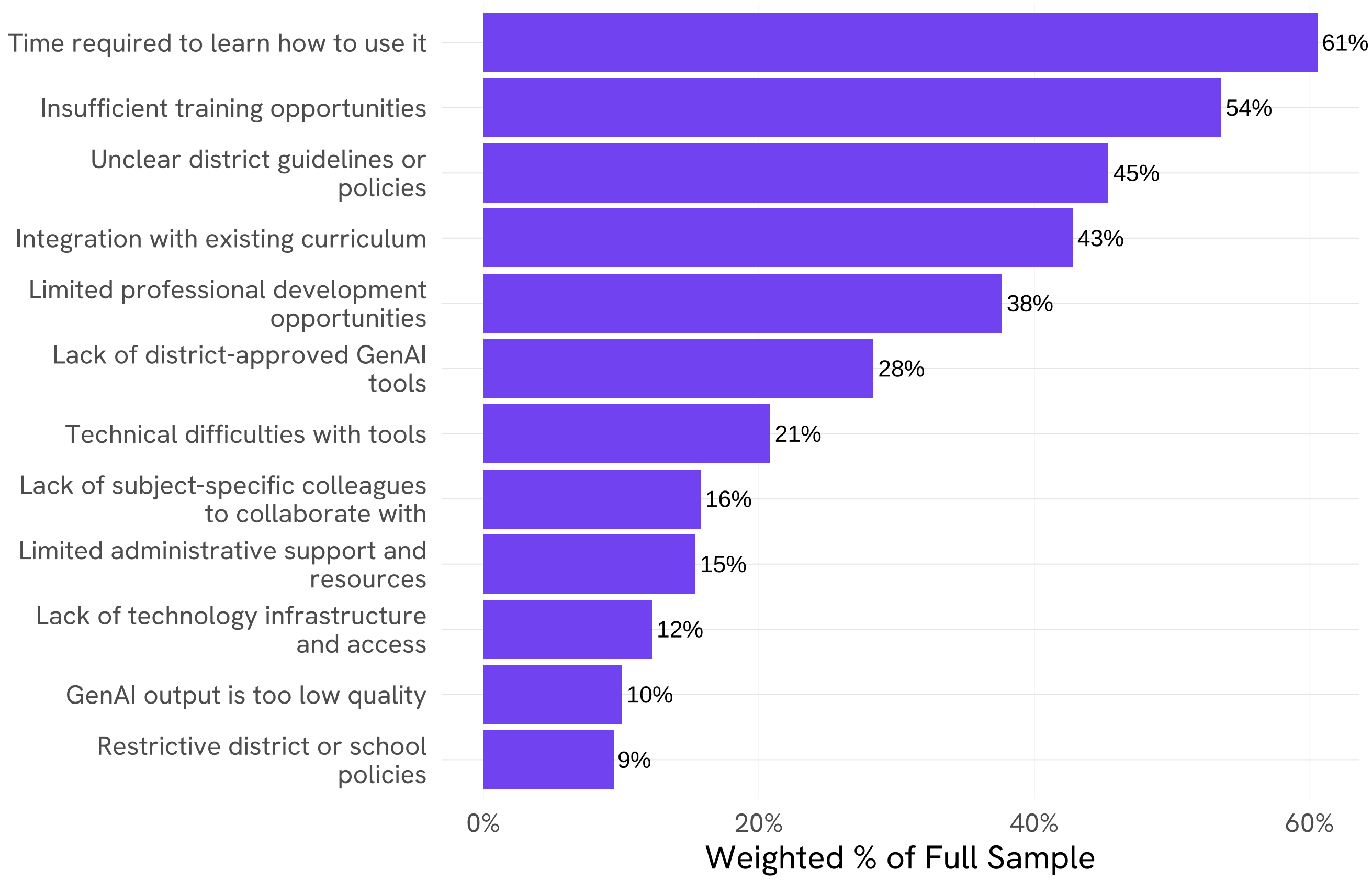

Question N = 979 (full sample)

**To what extent do you find the following to be concerns when using or considering GenAI in your teaching?**

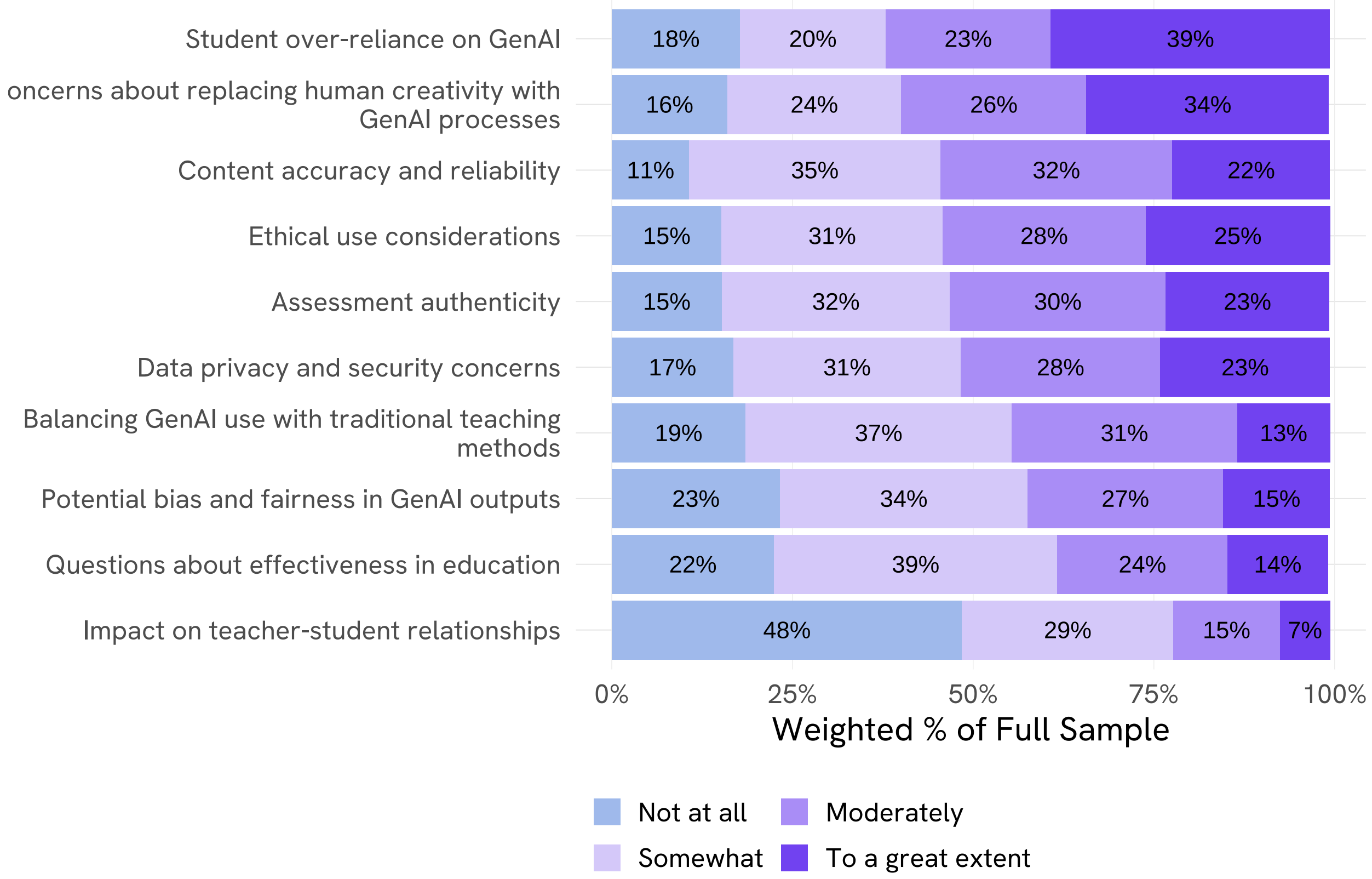

Question N = 979 (full sample)

# Discussion & Implications

This study provides a valuable glimpse into math and science teachers' adoption and use of generative AI for instructional tasks. It also provides an understanding of the policy and professional learning support required for GenAI to reach its potential as an instructional technology. 50% of teachers surveyed use GenAI in their teaching, and although only 44% of those use GenAI at least weekly, we anticipate adoption and use rates to increase, as 63% of GenAI users reported increasing their usage since they first started. The increased GenAI usage makes policy and professional learning support of growing importance.

Although finding efficiencies for overworked teachers is of value and can pay dividends as they have greater resources for the classroom, the potential of GenAI as an instructional technology lies in its potential for instructional improvement. 86% of GenAI-using teachers said they use it to be more efficient in their work, while a much smaller portion, between 20% and 46% used GenAI to actually improve their teaching by providing more relevant, real-world, cohesive learning experiences for students where they are asked better questions and receive better explanations and representations of concepts.

Why do we not have greater adoption of GenAI generally and specifically for the purpose of instructional improvement? We believe the answer lies in the still-nascent district support for GenAI adoption. For example, only 5% of survey respondents worked in schools with formal policy guidelines for teacher and student use, and only 9% of all teachers (17% of those reporting at least some district guidelines) said they worked in schools where GenAI integration into instruction was encouraged. Meanwhile, 45% of teachers said that a lack of clear district policy was a barrier to GenAI adoption, and the fact that 90% of teachers reported using ChatGPT, a GenAI model not built specifically for teaching and learning, much less mathematics and science, suggests that the lack of district policy extends to a lack of GenAI tools endorsed and supported as an instructional technology.

Importantly, teachers ranked assessment, feedback, and student progress tracking lowest on their list of ways GenAI has improved their workflows. Given the potential of GenAI to serve as an analytical tool for teachers to understand student thinking and learning and the profound data protection issues that accompany using GenAI with student data, we recommend that school and district leaders prioritize adopting tools with robust data protection agreements and provide professional development for teachers on GenAI assessment, feedback, and tracking uses.

Finally, teachers are worried about student use. They are concerned about how to help students learn to use GenAI as a learning partner instead of a learning substitute, with 34% indicating that GenAI has had negative impacts on student learning. 61% of teachers said the time to learn to use GenAI was a major barrier to adoption, with 54% reporting a lack of professional development as a barrier. Teachers want to learn about the technology and how to use it for lesson planning, differentiation, and assessment applications, but they also want professional development on how to help students understand the technology and use it appropriately.

We believe the findings from this survey serve as a clear call for greater resources devoted to math and science teachers' GenAI professional learning. We also believe that districts require their own learning support to understand GenAI as a technology and how to support teachers with GenAI tools and policies that help them achieve schools' instructional improvement goals. Further research is required to identify high-leverage policies and professional learning practices that schools and districts can employ to support teachers in deploying GenAI to improve teaching and learning.

# Technical Notes & References

## Survey Method and Sample Design

The 2025 Survey on Teachers' Use of Generative AI in Math and Science Instruction was conducted on behalf of the University of Washington by the RAND Corporation. The survey was administered to members of the American Teacher Panel (ATP), a nationally representative sample of approximately 25,000 K-12 public school teachers in the U.S. Teachers were recruited to the panel using probability-based sampling, ensuring broad demographic and geographic representation. The survey was conducted between April 30 and May 15, 2025, via online invitations and three reminder emails.

The survey targeted 1,000 completed responses. A total of 2,223 teachers were invited, and 1,016 teachers fully completed the survey, with 44 partial completions. 11 teachers were screened out as ineligible, and 14 additional responses were removed during data cleaning. The final analytic sample included 979 cases, with a completion rate of 44.5%.

The sample was intended to represent the national population of K-12th-grade public school math and science teachers. However, the survey under-represents Kindergarten teachers with only 26 collected responses. Survey weights were developed by RAND statisticians to account for:

- Probability of selection into the teacher panel.
- Likelihood of survey completion.
- Alignment with the national distribution of grade 1-12 teachers based on National Center for Education Statistics benchmarks (teacher-level data from the 2020-21 *National Teacher and Principal Survey*; school-level data from the 2022-23 *Common Core of Data*).

The final weights were trimmed (at the 95th percentile) and recalibrated to match NCES population statistics, ensuring accurate national representation in analysis.

Weighting aligned with key population variables, including:
- School level (elementary, middle, high)
- Locale (urban, suburban, rural)
- School size
- Minority percentage
- Poverty level
- Teacher gender, race/ethnicity, and experience

All reported percentages in this report are calculated using the survey weights. The margin of error is ±3.22% at 95% confidence for a binary variable with a 50% proportion, adjusted for weighting.

Additional technical documentation is available at [this link].

## Survey Instrument

This module of the survey included 20 closed-ended items and 3 open-ended items. Selected findings from open-text responses are included in this report alongside relevant charts to illustrate and contextualize quantitative findings. Formal qualitative coding of open-text responses is not included in this report. The full text of the survey questionnaire is available at this link.

The survey aimed to:

- Examine how, why, and to what extent math and science teachers are using generative AI (GenAI) in their instructional practices.
- Explore how GenAI adoption is reshaping instructional relationships, student engagement, and teaching effectiveness.
- Identify key barriers to GenAI adoption and the training/support teachers need for effective integration.

## Definition of GenAI

Before answering, participants were shown the following definition of GenAI:
*Generative Artificial Intelligence (GenAI) tools leverage data to detect patterns, generate data, automate tasks, and support decision-making. There are many types of GenAI tools that are currently in use to support classroom instruction, including chatbots (e.g., ChatGPT, Google Gemini, Anthropic Claude) and intelligent assistant systems (e.g., Colleague AI, MagicSchool, Khanmigo, Education Copilot).*

## Limitations

This survey captures a point in time - May of 2025 - and the perspectives of educators about their use and understanding of GenAI tools for teaching purposes. The availability and capabilities of GenAI software are changing rapidly. OpenAI announced 'study mode' on June 29th, and on June 30th, Google announced 30 free tools with Gemini in Classroom, to say nothing of the many other announcements from other technology providers. Although a definition of GenAI was provided at the start of the survey, teachers' own interpretations of GenAI are likely to vary in this rapidly changing area of technology development and may differ from the survey definition.

As with all self-reported survey data, results may be influenced by recall bias or social desirability bias. However, this survey was conducted independently without researcher supervision and anonymously, limiting the impact of these biases.

The analysis is limited to teachers of mathematics and science, including general elementary teachers who teach multiple subjects. With only 26 responses, Kindergarten teachers were relatively under-represented in the sample compared to other grade levels. Findings do not reflect the perspectives of teachers in other subject areas.

Closed-ended questions were developed to include a broad range of possible uses of and perspectives on GenAI in education. However, there may be additional views that were not represented by the allowed options. Future work should be conducted to qualitatively analyze open-response data from teachers.

For questions or further information about this report, please contact amplifylearn@uw.edu.